\documentclass[namedreferences]{solarphysics}
\usepackage[hyperref,optionalrh,solaromanenum]{spr-sola-addons} 
\usepackage{natbib}                     
\usepackage{graphicx}                    
\usepackage{color}                       
\usepackage{breakurl}                         

\chardef\us=`\_

\begin{document}

\begin{article}

\begin{opening}

\title{Optimal Fitting of the Freidberg Solution to {\em{In Situ}} Spacecraft  Measurements of Magnetic Clouds}

%
 \author[addressref={1},corref,email={qh0001@uah.edu}]{\inits{QH}\fnm{Qiang Hu}\lnm{}\orcid{0000-0002-7570-2301}}

%
\runningauthor{Hu, Q.}
\runningtitle{Optimal Fitting of 3D MC Model}

\address[id={1}]{Department of Space Science, and
Center for Space Plasma and Aeronomic Research (CSPAR),
The University of Alabama in Huntsville,
Huntsville, AL 35805, USA}

\begin{abstract}
We report, in detail, an optimization approach for fitting a three-dimensional (3D) magnetic cloud (MC) model to  {\em in situ} spacecraft measurements. The model, dubbed the Freidberg solution, encompasses 3D spatial variations in a generally cylindical geometry, as derived from a linear force-free formulation. The approach involves a least-squares minimization implemention with uncertainty estimates from magnetic field measurements. We present one case study of the MC event on 22 May 2007 to illustrate the method and demonstrate the satisfying result of the minimum reduced $\chi^2\lesssim 1$, obtained from the STEREO B (STB) spacecraft measurements. In addition, since the ACE spacecraft at Earth crossed the STB solution domain with an appropriate separation distance, the result from the optimally fitted Freidberg solution along the ACE spacecraft path is compared with the actual measurements of magnetic field components. A correlation coefficient of 0.89 is obtained between the two sets of data. 
\end{abstract}

%

\end{opening}

%
\section{Introduction}\label{sec:intro}
Magnetic clouds (MCs) are an important type of space plasma structures in the solar wind. They have been originally identified from {\em in situ} spacecraft measurements and possess a clearly defined set of observational signatures \citep[e.g.][]{1995ISAA....3.....B}. These include 1) enhanced magnetic field magnitude, 2) smooth rotation of one or more field components, and 3) depressed proton temperature or proton $\beta$ (the ratio between the plasma and magnetic pressure). They are believed to be interplanetary counterparts of coronal mass ejections (CMEs), although they may constitute a fraction of all interplanetary CMEs. These distinctive observational signatures, especially regarding the magnetic field, have led to the modeling of these structures by the configuration of a magnetic flux rope, often in a cylindrical geometry with one-dimensional (1D) or two-dimensional (2D) spatial symmetry.

A classical model for a 1D flux rope configuration is derived from the so-called Lundquist solution \citep{lund}, describing a 1D (with spatial dependence on the distance from a central axis only) cylindrical magnetic field structure in a force-free state (i.e., the Lorentz force vanishes), applicable to MCs (owing to the fact of low $\beta$). {\textbf{A number of studies has carried out Lundquist model fitting to {\em{in situ}} MC measurements \citep[e.g.,][]{JA093iA07p07217,Lepping1990,Lepping1997} for decades. Different models and variations were also devised and applied, such as the constant twist Gold-Hoyle model \citep{1999AIPCF,2003JGRA..108.1362D}, non-circular cross sections and/or toroidal geometry \citep{2017SoPh..292..129V,2012ApJ...748..109H,2003A&A...398..801V}, and non-force free models \citep[e.g.,][]{2002JGRA..107.1002H,2016ApJ...823...27N}, to name a few. They have made significant contributions to the analysis of MCs.  }} A unique and completely 2D flux rope model was developed based on the Grad-Shafranov (GS) equation governing a magnetohydrostatic equilibrium that goes beyond both the force-free condition and the 1D spatial geometry. In other words, the cross section has 2D variations, whereas the overall geometry remains cylindrical (or toroidal) with ignorable variation along the cylindrical (or the major) axis. The GS reconstruction method has been applied to Sun-Earth connection event studies for the past two decades \citep[see][for a recent review]{Hu2017GSreview}. The latest development includes an extension of the GS method to a toroidal geometry, while remaining 2D in nature \citep{2017Husolphys,2017SoPh..292..171H}.

{\textbf{It is desirable to extend the existing models to three-dimensional (3D) configurations for {\em in situ} spacecraft data analysis of MCs to account for well-known variabilities in such measurements. }} The latest attempt was made by \citet{2020HU3DMCGRL} and showed promising initial success. The viable approach is to exploit the existing  linear force-free field (LFFF) formulation. Specifically, as shown by \citet{freidberg}, the equation governing an LFFF $\mathbf{B}$ is a Helmholtz equation, $\nabla^2\mathbf{B}+\mu^2\mathbf{B}=0$, where the LFFF constant is denoted $\mu$. \textbf{To comply with the usual notion in the space physics community for the so-called constant $\alpha$ LFFF, we replace the symbol $\mu$ by $\alpha$ in this presentation. } In a cylindrical coordinates $(r,\theta, z)$, a set of series solutions in terms of the corresponding eigenfunction expansions is obtained by the standard approach of separation of variables. The eigenfunctions for the $r$ dimension are given by the Bessel's functions of the first kind, $J_n$, of increasing integral orders, $n=0,1,2,...$. By truncating the series solutions at $n=0$ for the $B_z$ component, the Lundquist solution is obtained, which only has $r$ dependence (thus completely 1D). If one includes higher-order modes, then the series solutions containing additional terms yield a more general solution which allows for additional variations in both $\theta$ and $z$ dimensions. Such a solution was presented in \citet{freidberg} and utilized in \citet{2020HU3DMCGRL}, dubbed the Freidberg solution (to be presented in Section~\ref{sec:approach} as well). The Freidberg solution is intrinsically 3D, although the variation in the $z$ dimension is periodic with a finite wavelength. 

The focus of this report is to describe in detail the practical procedures developed for an optimal fitting of the Freidberg solution to {\em in situ} spacecraft measurements of MCs. We further demonstrate the merit of this approach by one case study of a rare MC event observed by two spacecraft simultaneously. A correlative analysis is thus enabled by this occurrence and serves as a unique opportunity for validation of this approach. The recipe for the fitting procedures is given in Section~\ref{sec:approach}. The two-spacecraft case study is presented in Section~\ref{sec:validation}. The last section summarizes the results. 

\section{Optimal Fitting Procedures}\label{sec:approach}
The optimal fitting algorithm we employ follows the standard procedures for a least-squares minimization problem described in \citet{2002nrca.book.....P}. Namely, an objective function is expressed as an evaluation of the square difference between a set of vector magnetic field measurements $\mathbf{b}$ and the corresponding set $\mathbf{B}$ yielded by an underlying model
\begin{equation}
\chi^2=\frac{1}{\tt{dof}}\sum_{\nu=X,Y,Z}\sum_{i=1}^N\frac{(b_{\nu i} - B_{\nu
i})^2}{\sigma_{ i}^2}. \label{eq:chi2}
\end{equation}
\textbf{The $\chi^2$ value is  usually normalized by the degree-of-freedom, ${\tt{dof}}=3N-p-1$, with $N$ the number of magnetic field vectors and $p$ the number of to-be-determined parameters among the set, resulting in the reduced $\chi^2$ value. The associated uncertainties in the measured magnetic field  are denoted $\sigma_i$. }

The underlying model is provided by the Freidberg solution \citep{freidberg}, as given below in a cylindrical coordinate system $(r,\theta,z)$,
\begin{eqnarray}
\frac{B_z(\mathbf{r})}{B_{z0}} & = & J_0(\alpha r)+CJ_1(l r)\cos(\theta+kz) \label{eq:B}\\
\frac{B_\theta(\mathbf{r})}{B_{z0}} & = & J_1(\alpha r)-\frac{C}{l}\left[\alpha J'_1(l r)+\frac{k}{l r}J_1(l r)\right]\cos(\theta+kz) \\
\frac{B_r(\mathbf{r})}{B_{z0}} & = & -\frac{C}{l}\left[k J'_1(l r)+\frac{\alpha}{l r}J_1(l r)\right]\sin(\theta+kz). \label{eq:B4}
\end{eqnarray}
The objective function is to be minimized, yielding the set of parameters for the corresponding Freidberg solution, subject to the set of spacecraft measurements of magnetic field $\mathbf{b}$ along the spacecraft path, and the associated uncertainties $\sigma_i$. \textbf{The  data points for analysis $\mathbf{b}_i$  are averaged from higher-resolution (e.g., 1-miniute) magnetic field data $\tilde\mathbf{B}$ over each hourly interval, for instance, for an MC interval of $N$ hours, i.e., $\mathbf{b}_i=\langle\tilde\mathbf{B}\rangle_{(i\mathrm{th}~\mathrm{hour})}$ ($i\in[1,N]$). The associated uncertainties are estimated by the corresponding root-mean-square (RMS) values of the higher-resolution magnetic field components, i.e., $\sigma^2_i=\sum_\nu\langle(\tilde{B}_\nu-\langle\tilde{B}_\nu\rangle)^2\rangle_i$.   }

The set of parameters includes $B_{z0}$, $C$, $\alpha$, and $k$ (note that $l=\sqrt{\alpha^2-k^2}$). {\textbf{Generally the parameter $B_{z0}$ is pre-determined as the average magnetic field magnitude from the spacecraft measurements.}} In addition, the geometrical parameters include the directional angles of the $z$ axis orientation in space, $(\delta,\phi)$, the translation along $z$, $z_{\min}$, the rotation around the $z$ axis, $\theta_{\min}$, and the translation of the $z$ axis on the $(x,y)$ plane, $(X_0,Y_0)$. They need to be considered to account for the arbitrary geometry of the solution domain, relative to the spacecraft path, when working in a frame co-moving with the structure to be fitted with the Freidberg solution.
 The reduced $\chi^2$  is the value we calculate and present throughout the procedures, as given by equation~(\ref{eq:chi2}). 

The procedures start by enumerating all possible $z$ axis orientations in space by going through pairs of the angles $(\delta,\phi)$ over a grid with uniform spacing $6^{\circ}\times 6^\circ$. For each pair, the corresponding trial $x$ and $y$ axes are also chosen to form a right-handed coordinate system. This step is necessary for reducing the number of unknown parameters to be determined and serves the purpose of choosing ``well-educated'' initial guesses for the final optimization steps as to be described below. As mentioned earlier, for practical applications, we allow the solution domain to translate both along and perpendicular to the $z$ axis, as well as to rotate around the $z$ axis. These operations are equivalent to adding free parameters, $z_{\min}$, and $\theta_{\min}$, explicitly to the dimensions $z$, and $\theta$, respectively, and $(X_0,Y_0)$ implicitly to $r$, in equations~(\ref{eq:B}-\ref{eq:B4}). These additional parameters will account for an arbitrary spacecraft path across the Freidberg solution domain due to the lack of symmetry, thus increasing the chances of yielding an optimal fit to the spacecraft measurements.

 \begin{figure} 
 \centerline{\includegraphics[width=0.5\textwidth,clip=]{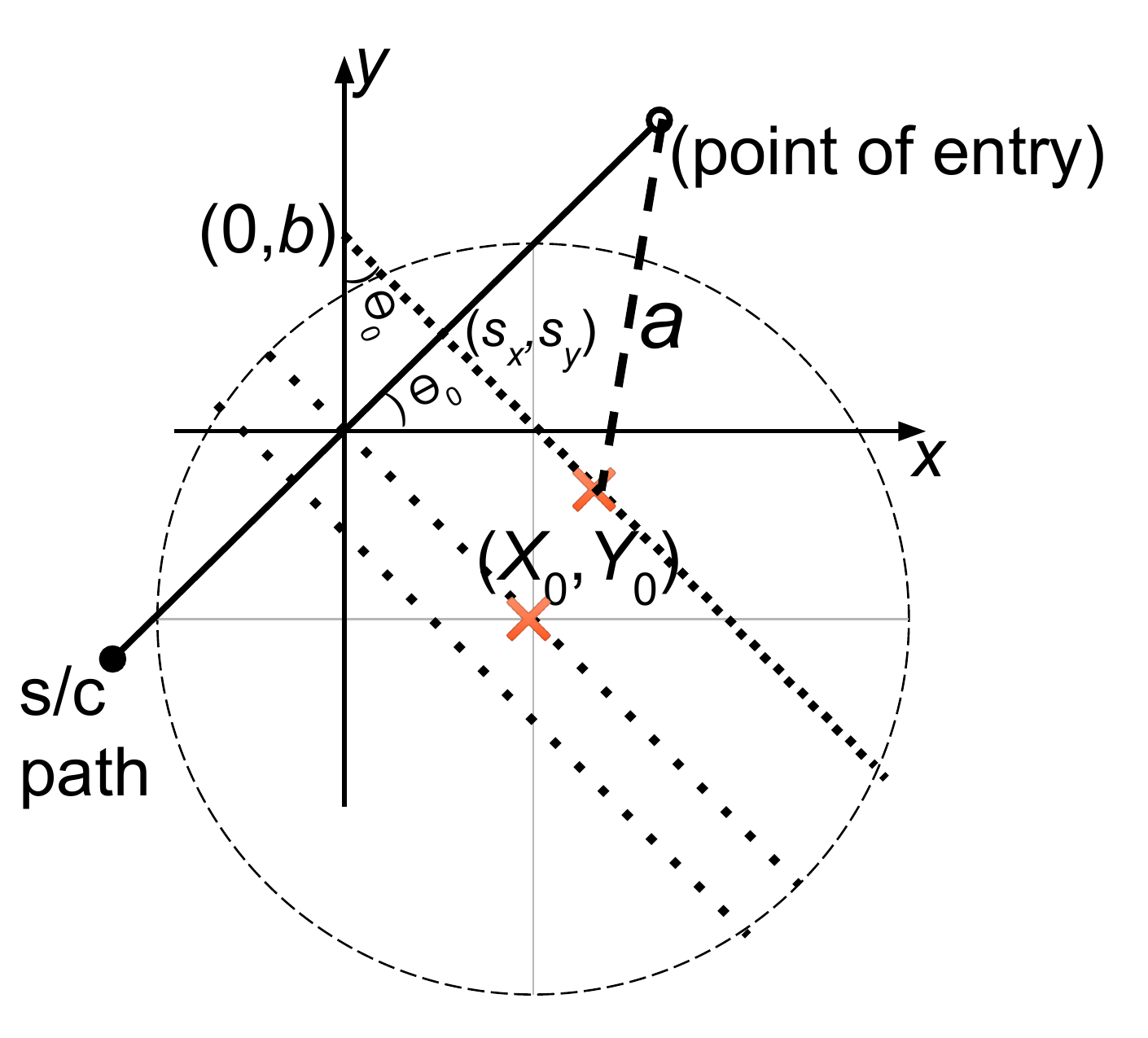}}
 \caption{The schematic for the coordniates $(x,y,z)$ and associated geometrical parameters as seen on the $(x,y)$ plane. See text for details. The crosses mark the locations of $(X_0,Y_0)$, along the parallel dotted lines \textbf{which are perpendicular to the projected spacecraft path. The distance between the point of entry and the origin is $a_0$. The dashed line illustrates the parameter $a$, length of a line segment connecting the ``point of entry" and the point  $(X_0,Y_0)$.}}\label{fig:coords}
 \end{figure}
An important parameter, $a$, associated with this geometry and also serving as the normalization constant for length scales, is to be worked out, as illustrated in Figure~\ref{fig:coords}.  Figure~\ref{fig:coords} shows the schematic of the $(x,y,z)$ coordinates as viewed down the positive $z$ axis onto the $(x,y)$ plane, and the associated parameters. For each trial $z$ axis, the spacecraft path can be projected onto the $(x,y)$ plane for a selected MC interval with duration $T$. The origin $(0,0)$ is chosen at a location along the  spacecraft path, \textbf{e.g., at the center of the MC interval. The projected spacecraft path length between the point of entry and the origin, $a_0$, is calculated beforehand from the spacecraft measurements.} The locations of $(X_0,Y_0)$ are set to be along a group of parallel lines perpendicular to the projected spacecraft path, as illustrated by the dotted lines in Figure~\ref{fig:coords}. They satisfy the equation $y=\frac{-1}{\tan\theta_0}x+b$, with an additional parameter $b$, while the angle $\theta_0$ is known for the chosen coordinates. These lines intercept the projected spacecraft path at the point $(s_x,s_y)$. Then the following equation is satisfied (\textbf{applying the pythagorean theorem for the right triangle with vertices, ``point of entry", $(X_0,Y_0)$, and $(s_x, s_y)$ as denoted in Figure~\ref{fig:coords}}), 
\begin{equation}
(X_0-s_x)^2a^2+(Y_0-s_y)^2a^2=a^2-(a_0-ba\sin\theta_0)^2\approx a^2-a_0^2. \label{eq:X0Y0}
\end{equation}
Hereafter the parameters (e.g., $X_0,Y_0$, and $b$) with spatial dimensions are normalized by $a$, except for $a_0$, \textbf{a known value determined at the beginning from spacecraft measurements (see below the detailed steps). Besides it may serve as an estimate of the cross-section size in physical unit of the structure. } It follows \textbf{from equation~(\ref{eq:X0Y0}) by omitting the term $ba\sin\theta_0$ in the right-hand side}
\begin{equation}
a=\frac{a_0}{\sqrt{1-[(X_0-s_x)^2+(Y_0-s_y)^2]}}, \label{eq:a}
\end{equation}
with
$s_x=b\sin\theta_0\cos\theta_0$, and $s_y=b\sin^2\theta_0$. Since the parameter $a$ mostly serves as a normalization constant, and carries no physical meaning, an accurate determination is not necessary. Therefore we apply an approximation in equation~(\ref{eq:X0Y0}) to avoid evaluating the roots to the quadratic equation in $a$. 

In practice, it is necessary to further reduce the number of free parameters to be determined, especially by employing a simple unbounded optimization procedure, namely, the function $\tt{fminsearch}$, implemented in Matlab.  Therefore we devise a multi-step minimization approach, beginning with an optimization procedure with the minimum number of free parameters. We start the optimal fitting procedures with pre-defined fixed values for $ka=1.2$ and $\alpha a=3.11$, which correspond to a solution of minimum magnetic energy state as shown by \citet{freidberg}. In the final steps, these restrictions are relaxed and they are allowed to vary and to be optimized. The detailed steps of the procedures are described below.
\begin{enumerate}
\item Select the data interval for analysis. Usually the data used include both magnetic field vectors  and bulk plasma parameters in the spacecraft centered Raidal, Tangential, and Normal (RTN) coordinates (preferred, but others work too). The plasma measurements are needed to assess the satisfaction of model assumptions and will assist in the selection of analysis interval. For instance, an interval may be chosen with clearly depressed proton $\beta$ value. The flow velocities are utilized to derive the deHoffmann-Teller (HT) frame velocity\citep{1998ISSIRK}, $\mathbf{V}_{HT}$, with which the MC structure is moving past the stationary spacecraft. Conversely, in the HT frame in which the structure appears stationary, the spacecraft is traversing the structure with a constant velocity, $-\mathbf{V}_{HT}$. Then the path length along the spacecraft path for the chosen  MC interval with duration $T$ is $L_{sc}=|\mathbf{V}_{HT}|T$. \textbf{The plasma measurements including both proton number density and velocity are needed to perform the HT analysis (see Appendix~\ref{sec:app} for the determination of and the justification for the HT frame).}
\item Set up a grid for the directional angles of the trial $z$ axis by using two angles, $\delta\in[0,180^\circ]$, and $\phi\in[0,360^\circ]$. Once a trial $z$ axis is chosen, set up the corresponding $x$ and $y$ axes in terms of $(\delta,\phi)$ as well. For each trial $(x,y,z)$ coordinate system, as illustrated in Figure~\ref{fig:coords}, project the spacecraft path onto the $(x,y)$ plane, and obtain the projected spacecraft path length $L_0$ (known from $L_{sc}$) and set $a_0=L_0/2$. 
According to Figure~\ref{fig:coords}, obtain $\theta_0$ and the coordinates of the point of entry, $S_0=[\cos\theta_0,\sin\theta_0,0]a_0$. Then subsequently the $(x,y,z)$ coordinates of a set of points along the spacecraft path are obtained, which are used for the evaluation of $\mathbf{B}$ from the Freidberg solution. We first carry out the $\chi^2$ minimization by using only three free parameters, $C$, $\theta_{\min}$, and $z_{\min}$, while looping through a discrete set of values for $b\in [-0.5,0.5]$, for example. In addition, a discrete set of $X_0\in[-1,1]$ is chosen, so that $Y_0=(-\frac{1}{\tan\theta_0})X_0+b$ is also known. In effect, there are two ``$\tt{For}$'' loops for iterations over the sets of $X_0$ and $b$ values, respectively. In the innermost loop, an optimization procedure is carried out to yield the corresponding $\chi^2$ value, and the associated set of parameters, $C$, $\theta_{\min}$, and $z_{\min}$, which are  sorted. Then the minimum $\chi^2$ value, $\chi^2_{\min}$, and the associated parameters are recorded. 
\item For each trial $z$ axis, i.e.,  at the end of  each loop over a pair of $(\delta,\phi)$, an additional optimization procedure is performed by relaxing the parameters $k$ and $\alpha$ \textbf{(no longer fixed), i.e., allowing them to vary freely together with $C$, $\theta_{\min}$, and $z_{\min}$}, in order to further reduce the minimum $\chi^2$ value.
\item When the iteration over $(\delta,\phi)$ is completed, the corresponding $\chi^2_{\min}$ and the associated parameters, $\delta$, $\phi$, $X_0$, $Y_0$, $b$, $C$, $\theta_{\min}$, $z_{\min}$, $k$, and $\alpha$ are obtained. The last step is to search for an optimal solution by further allowing the preset parameters to vary freely, especially for the pair $(\delta,\phi)$, with the values obtained at the end of Step iii) as ``good'' initial guesses. \textbf{No more iterations are needed. This can be achieved in several ways by progressively adding additional  parameters, $b$, $X_0$, and $Y_0$, one by one,  to the argument list of the optimization function, $\tt{fminsearch}$, as free parameters.} The final optimal solution is obtained from the output that yields the minimum $\chi^2$ value among the multiple optimization runs. The solution is deemed acceptable, when the associated value $Q= 1 - \tt{chi2cdf}(\chi^2, \tt{dof})$, where the function $\tt{chi2cdf}$ is the cumulative distribution
function of $\chi^2$,  significantly exceeds $10^{-3}$ \citep[see, e.g.,][]{2002nrca.book.....P}. 

\end{enumerate}
\textbf{We intend to provide a detailed descriptions of the working recipe here so that interested readers can develop their own implementations. To facilitate this effort, we are willing to provide the computer codes based on the above procedures upon request. }

\section{Method Validation: Two-Spacecraft Measurements}\label{sec:validation}
%
 \begin{table}
 \caption{List of selected parameters for the magnetic field configuration of the MC on 22 May 2007.}\label{tbl:para}
 \begin{tabular}{ccc}     
 \hline
 Parameter & Freidberg Solution & GS Result\tabnote{From M\"{o}stl et al., (2009a).}\\
 \hline
 $B_{z0}$ (nT) &		15.9					& 16.2 \\
 $D\approx 2a_0$ (au)        &    0.137       &  0.123 \\
 $C$              &    -0.886         & ...  \\
 $ka$              &     1.13           & 0  \tabnote{For a 2D configuration.} \\
 $\alpha a$         &      2.66          & ... \\  
 Chirality & +           & + \\
 $\hat{\mathbf z}$ in $(\theta,\phi)$\tabnote{The inclination angle $\theta$ to the ecliptic, and the azimuthal angle $\phi$ measured from -R towards -T axes, all in degrees.} &    $(34,73)$        & $(55,75)$ \\
 $\Phi_t$ ($10^{20}$Mx)  &   2.6          &  3.3$\pm 0.5$\\
 \hline
 \end{tabular}
 \end{table}
  
  We have shown in \citet{2020HU3DMCGRL} the initial application of the optimal fitting approach to {\em in situ} spacecraft measurements by the Freidberg solution. The merit of the results was justified by the criteria such that the minimum reduced $\chi^2\approx 1$ and $Q\gg 10^{-3}$, based on $\chi^2$ statistics \citep{2002nrca.book.....P}. To further validate the approach for practical applications, we present an additional case study here by using {\em in situ} measurements from two spacecraft with an appropriate separation distance. This is a rare occurrence during the early stage of the twin Solar and TErrestrial RElations Observatory
 (STEREO) spacecraft mission. On 22 May 2007, the two STEREO spacecraft, Ahead and Behind (hereafter STA and STB, respectively),  were separated from Earth by $\sim 6^\circ$ and $\sim 3^\circ$, respectively, in the ecliptic plane, with STA leading Earth to the west and STB lagging behind to the east. Both STB and the Advanced Composition Explorer (ACE; at Earth) spacecraft observed an MC event with distinct and similar signatures, based on {\em in situ} measurements, while the signatures at STA for an encounter of the same MC structure are much less clear. This event has been studied extensively, by a number of prior studies, employing multiple spacecraft measurements \citep{2008ApJ...677L.133L,2009JGRAM,2009SoPh..254..325K,2010JA015552}, inter-relating the observations of the same event at different locations and to the solar sources. In particular, \citet{2009JGRAM} performed a comprehensive study of this event and reconstructed a 2D configuration by combining the GS reconstruction results from both STB and Wind spacecraft measurements \citep[see, also,][]{2008AnGeo..26.3139M}. We compare a selected number of parameters characterizing the magnetic field configurations between our results and those from \citet{2009JGRAM}, as listed in Table~\ref{tbl:para}.

 \begin{figure} 
 \centerline{\includegraphics[width=0.3\textwidth,clip=]{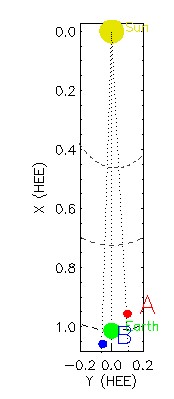}
 \includegraphics[width=0.66\textwidth,clip=]{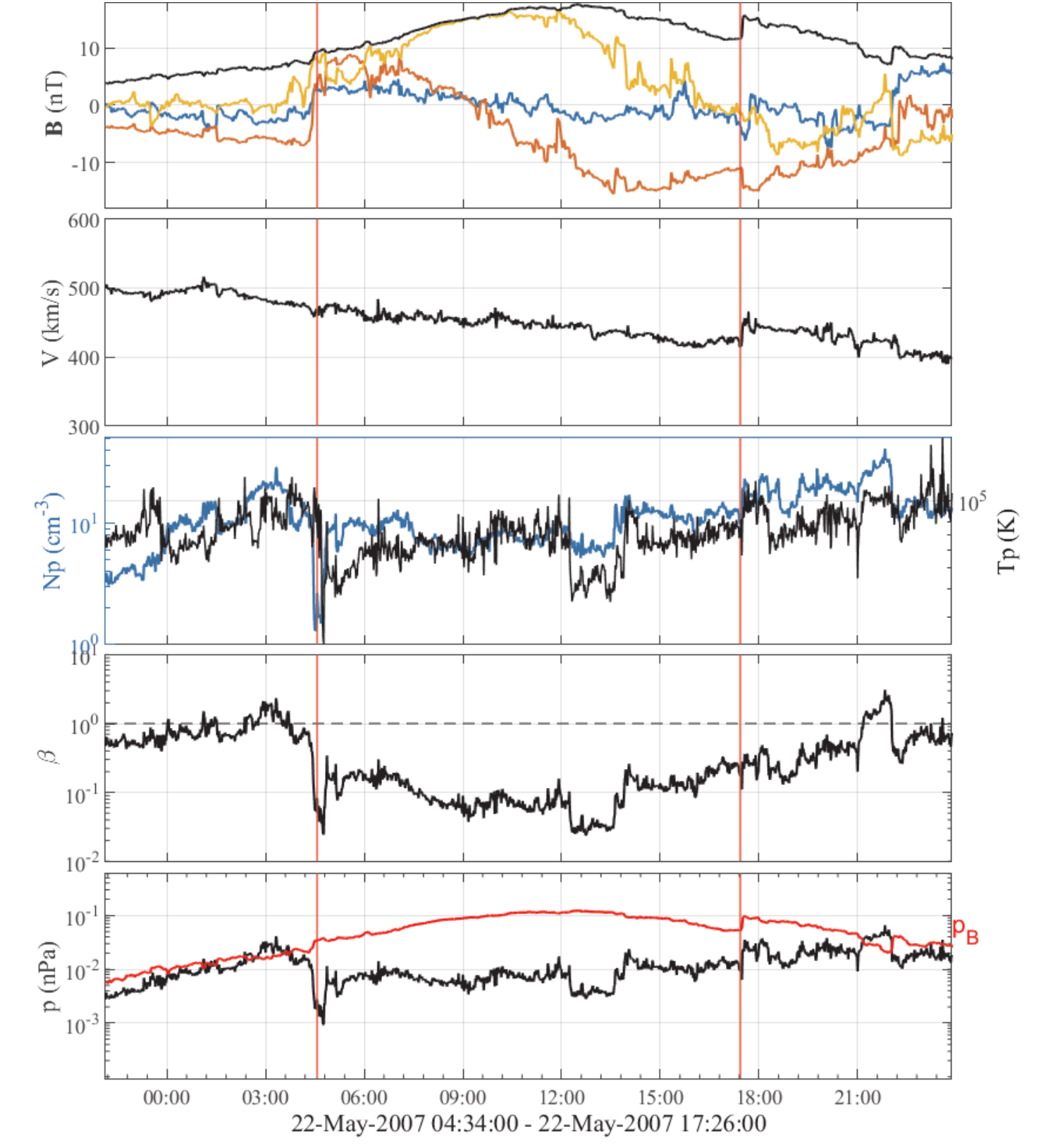}}
 \caption{Left panel: the spacecraft locations of STA (A), STB (B), and ACE (Earth) on 22 May 2007 on the ecliptic plane in the Earth Ecliptic coordinates (HEE; courtesy of the STEREO Science Center); Right panels: the time series data from STB spacecraft measurements with 1-min cadence. In the right, the top to bottom panels are, the magnetic field  in R (blue), T (brown), N (gold) components and its magnitude (black), the solar wind speed, the proton number density (left axis) and temperature (right axis), the proton $\beta$, and the plasma pressure (black) and the magnetic pressure (red). The vertical lines mark the interval selected for analysis as indicated beneath the last panel.}\label{fig:STBdata}
 \end{figure}
 
\begin{figure} 
 \centerline{\includegraphics[width=0.5\textwidth,clip=]{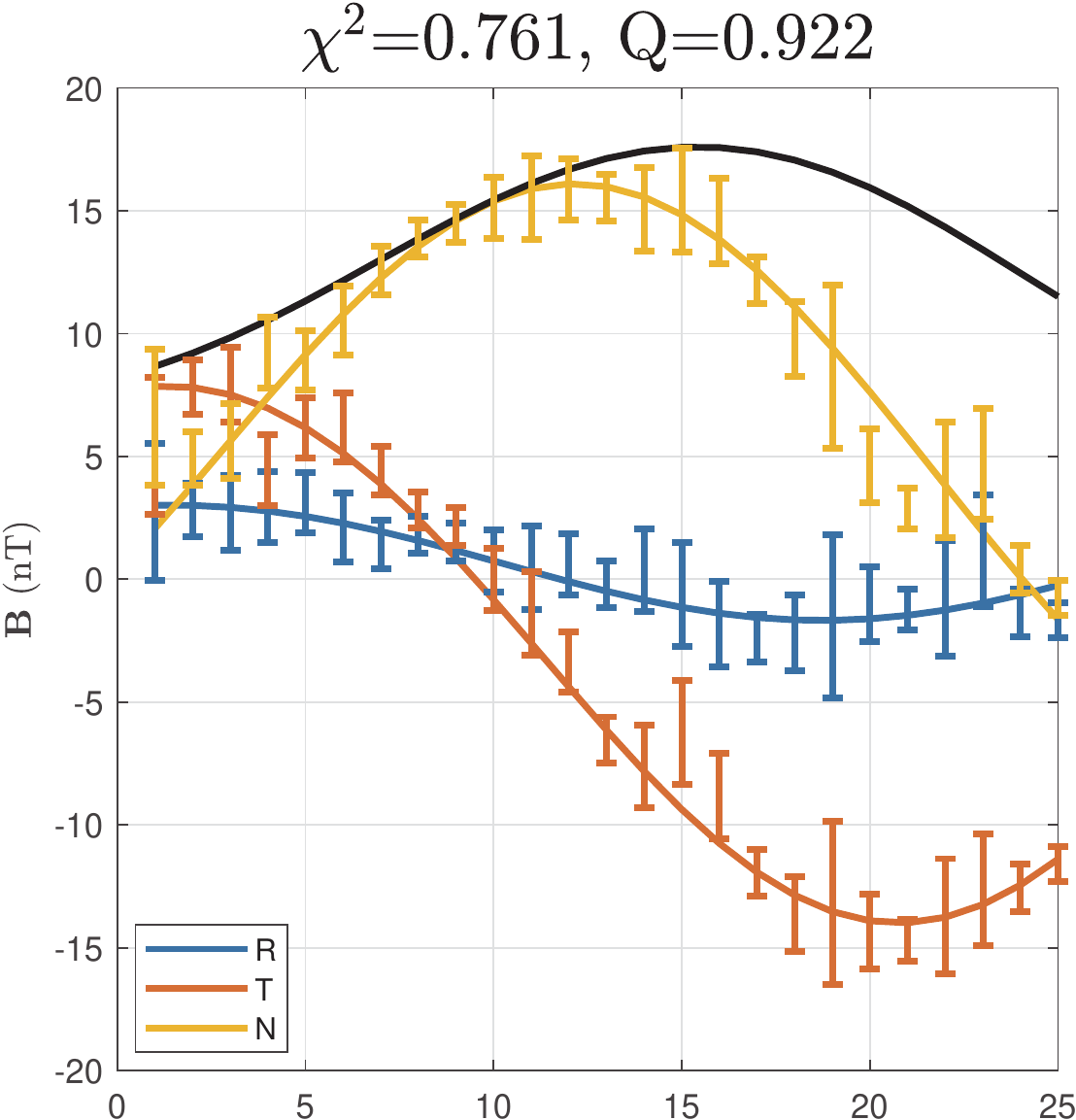}}
 \caption{The optimal fitting result for the selected STB MC interval. The solid curves are the output from the optimal Freidberg solution. The errorbars are the STB measurements averaged to 25 points across the MC interval with uncertainty estimates from the RMS values of the underlying 1-min resolution data.  The black curve represents the field magnitude. The corresponding minimum reduced $\chi^2$  and $Q$ values are denoted on top.}\label{fig:Brtn}
 \end{figure}
\begin{figure} 
 \centerline{\includegraphics[width=0.5\textwidth,clip=]{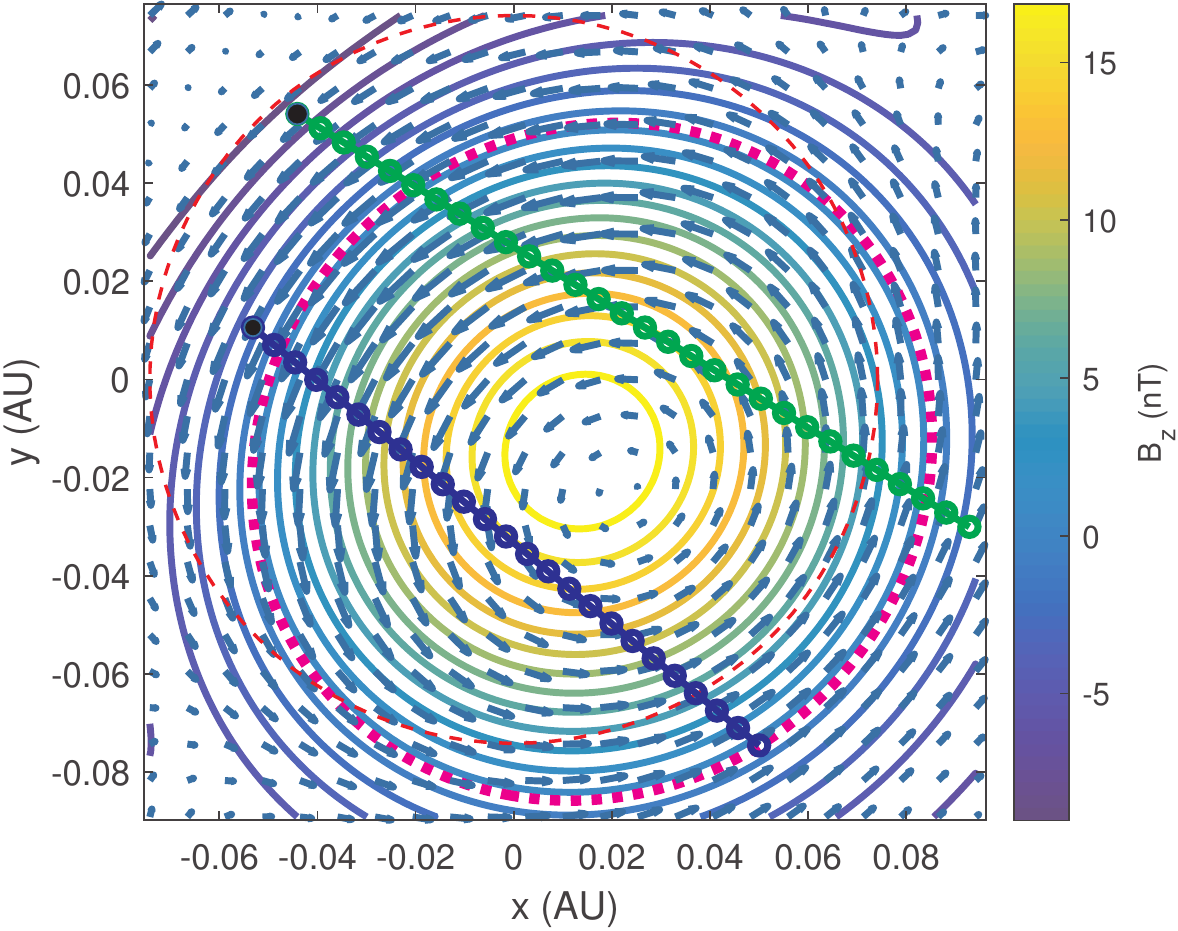}}
 \caption{The cross section of the magnetic field configuation at $z=0$. The contours are the axial field $B_z$ with scales indicated by the colorbar. The arrows represent the corresponding transverse field vector $(B_x,B_y)$. The thin dashed red curve represents the circle $r=a$, while the thick dotted magneta curve marks the boundary $B_z=0$. The blue and green dots forming straight lines depict the STB and ACE spacecraft paths, respectively. }\label{fig:Bz0}
 \end{figure}
\begin{figure} 
 \centerline{\includegraphics[width=0.4\textwidth,clip=]{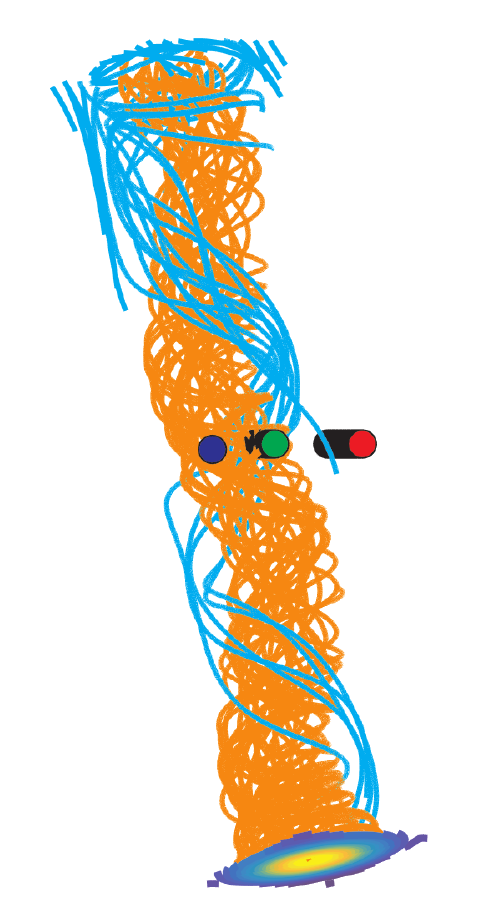}}
 \caption{A 3D view of magnetic field lines along the STB spacecraft path (blue dots) toward the Sun. The ecliptic north is upward. The green (red) dots to the west mark the ACE (STA) spacecraft path in this view. The orange lines originate from the positive polarity region, while the cyan lines end in the negative polarity region, on the bottom plane.}\label{fig:fldlines}
 \end{figure}
Figure~\ref{fig:STBdata} shows the STB {\em in situ} measurements of the MC event with the analysis interval marked for the optimal fitting  by the Freidberg solution, following the procedures described in Section~\ref{sec:approach}. The proton $\beta$ is low throughout the interval, $\langle\beta\rangle\approx 0.11$, approximately satisfying the force-free assumption.  An optimal solution is achieved as shown in Figure~\ref{fig:Brtn}, with the corresponding fitting parameters and derived physical quantities listed in Table~\ref{tbl:para}. The geometrical parameters compare well with the corresponding GS reconstruction result, in terms of the overall cross-section size $D$, and the orientations of $\hat{\mathbf z}$. Both yield a flux rope configuration with the right-handed chirality. The axial field and the amount of toroidal (axial) flux $\Phi_t$ are consistent. However the Freidberg solution exhibits distinctive 3D features, as to be demonstrated below, which are not present in 2D configurations.

A cross section plot is shown in Figure~\ref{fig:Bz0}, where the two spacecraft paths of STB and ACE, as viewed down the $z$ axis, are depicted. A boundary may be defined as a surface where $B_z=0$, as illustrated here by the closed magenta dotted line on the $z=0$ plane. \textbf{Note that the configuration as shown changes with $z$.} The two spacecraft paths were separated by $\sim$0.05 au in distance, although each path is not equivalent to lying on one  cross-section plane, owing to the nature of a 3D configuration. This feature is better seen via a 3D view of the field-line configuration, as shown in Figure~\ref{fig:fldlines}, from the perpective of STB toward the Sun. Two sets of field lines are drawn: one in orange represents the main flux bundle winding upward (approximately northward), rooted on the major positive polarity region (enclosed by the boundary $B_z=0$), while the cyan lines are the (minor) flux bundle winding downward. Two bundles are wrapping around and end at neighboring positive and negative polarity regions, respectively. Because they both possess the right-handed chirality (positive $\alpha$ value; see Table~\ref{tbl:para} and Figure~\ref{fig:Bz0}), the field is continuous between the interface (where $B_z=0$) of the two flux bundles.  Again, similar to Case 1 presented in \citet{2020HU3DMCGRL}, we obtain a more general 3D flux rope configuration, exhibiting a winding body of flux bundles with mixed polarities of varying strengths. {\textbf{The major distinction from a Lundquist type solution, which also exhibits mixed polarities but is largely 1D, is that the spatial variation in a Freidberg solution is 3D, which yields a greater deal of variations in the configuration of the mixed polarity regions.} Here the positive polarity dominates (but see  Case 2 in \citet{2020HU3DMCGRL} for the solution of equally mixed polarities/helical states where they are located side by side).} Such a flux rope configuration is distinctive from a cylindrical configuration with an identifiable straight central field line, generally indicative of negligible variation along the $z$ axis.

 \begin{figure} 
 \centerline{\includegraphics[width=0.5\textwidth,clip=]{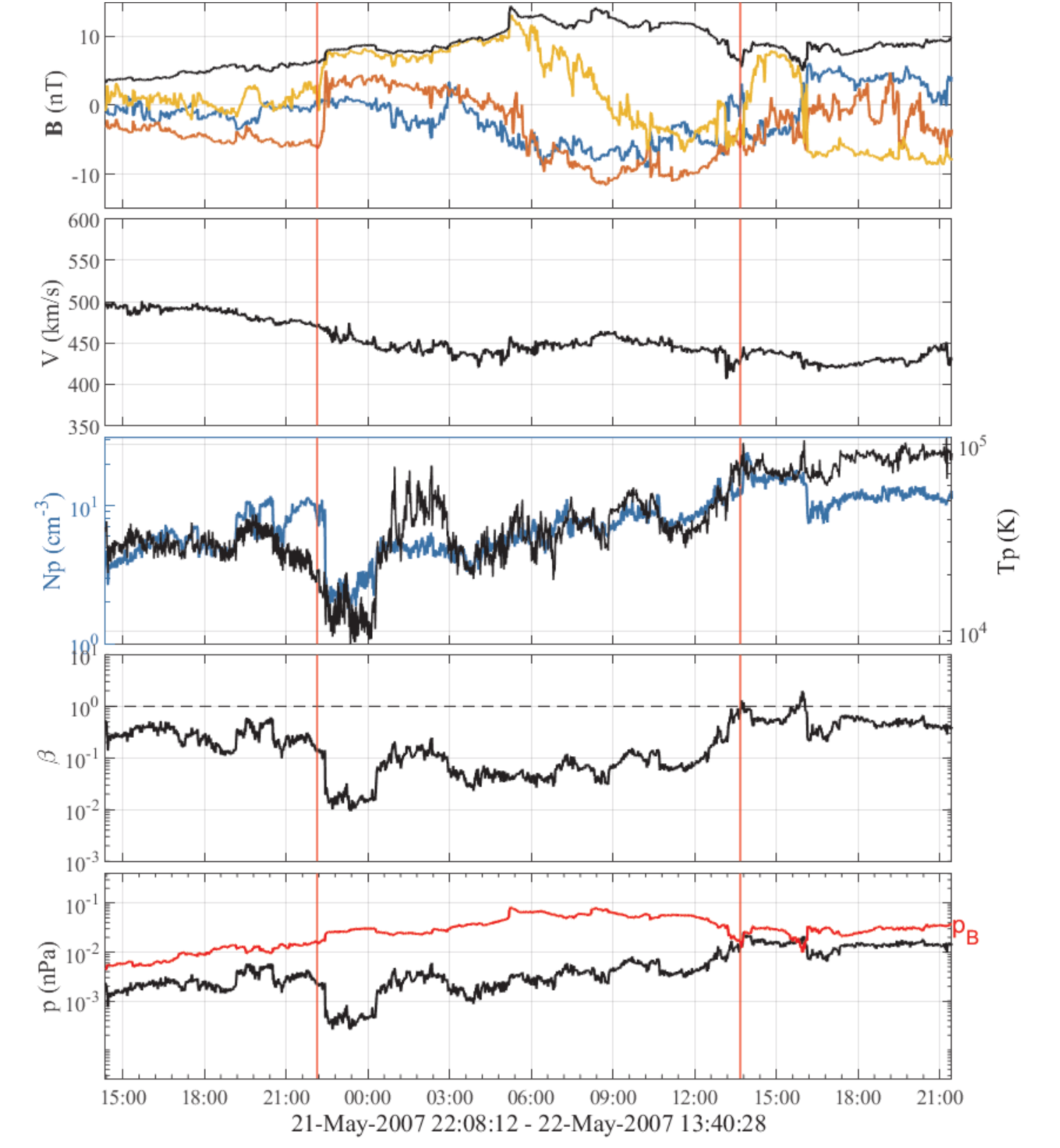}
 \includegraphics[width=0.5\textwidth,clip=]{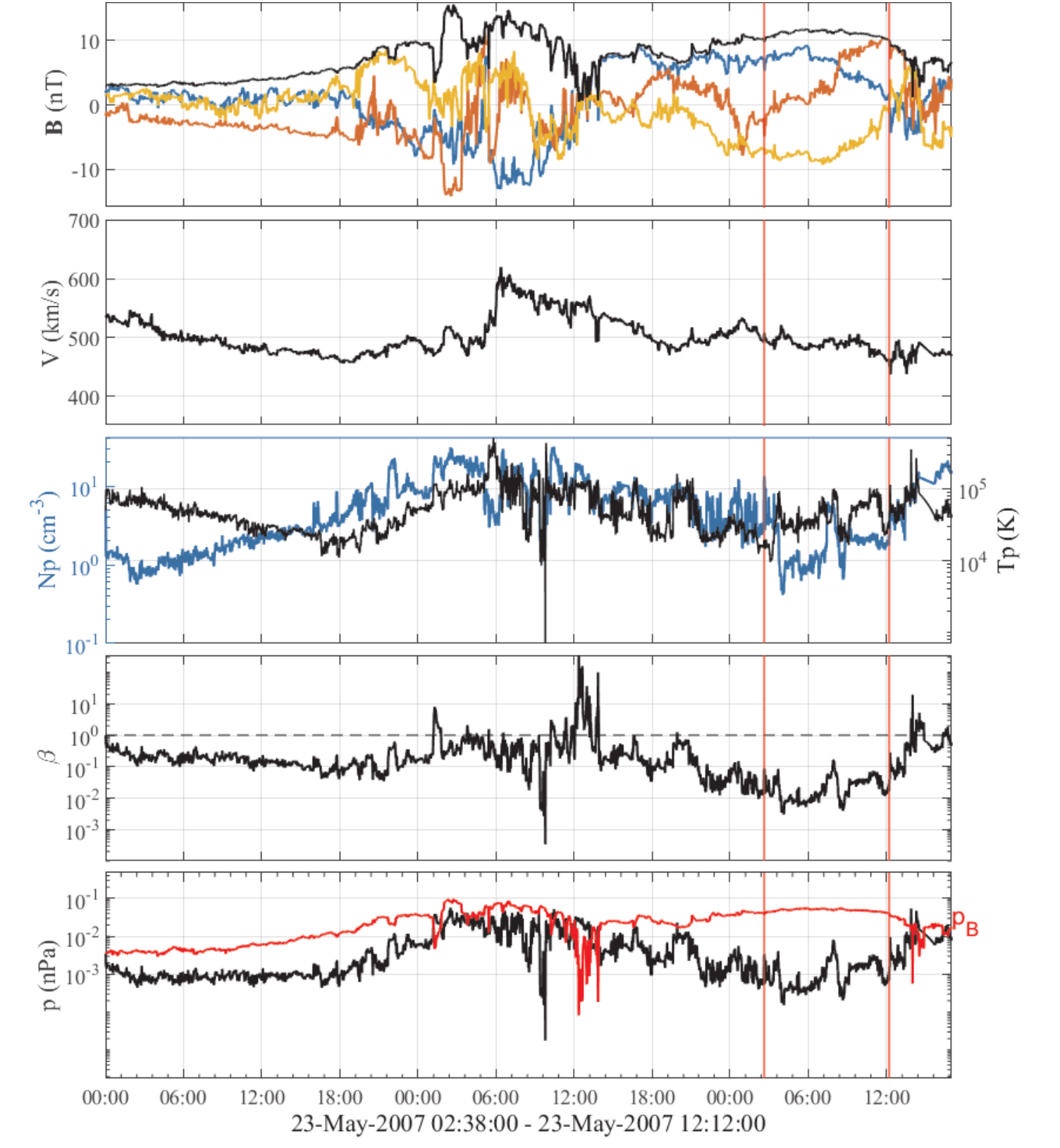}}
 \caption{\textbf{The ACE spacecraft {\em{in situ}} measurements (left panels), and the STA spacecraft {\em{in situ}} measurements (right panels; starting time: 00:00 UT 21 May 2007). Format is the same as Figure~\ref{fig:STBdata}. In  the left panels, the vertical lines mark the selected  interval from the ACE dataset for comparison purpose only. The MC interval marked in the right panels on 23 May 2007 is taken for an optimal fitting to the Freidberg solution.}}\label{fig:ACEdata}
 \end{figure}

\begin{figure} 
 \centerline{\includegraphics[width=0.55\textwidth,clip=]{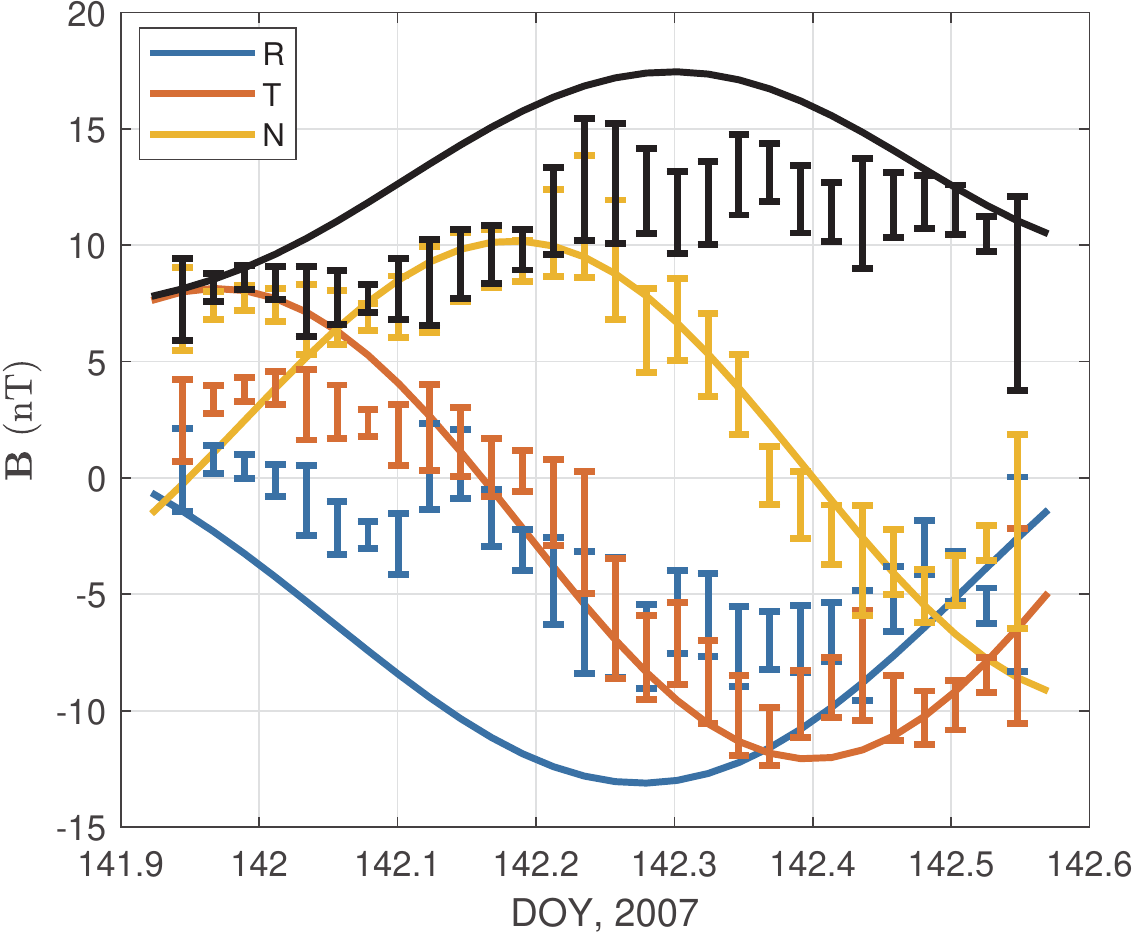}
  \includegraphics[width=0.45\textwidth,clip=]{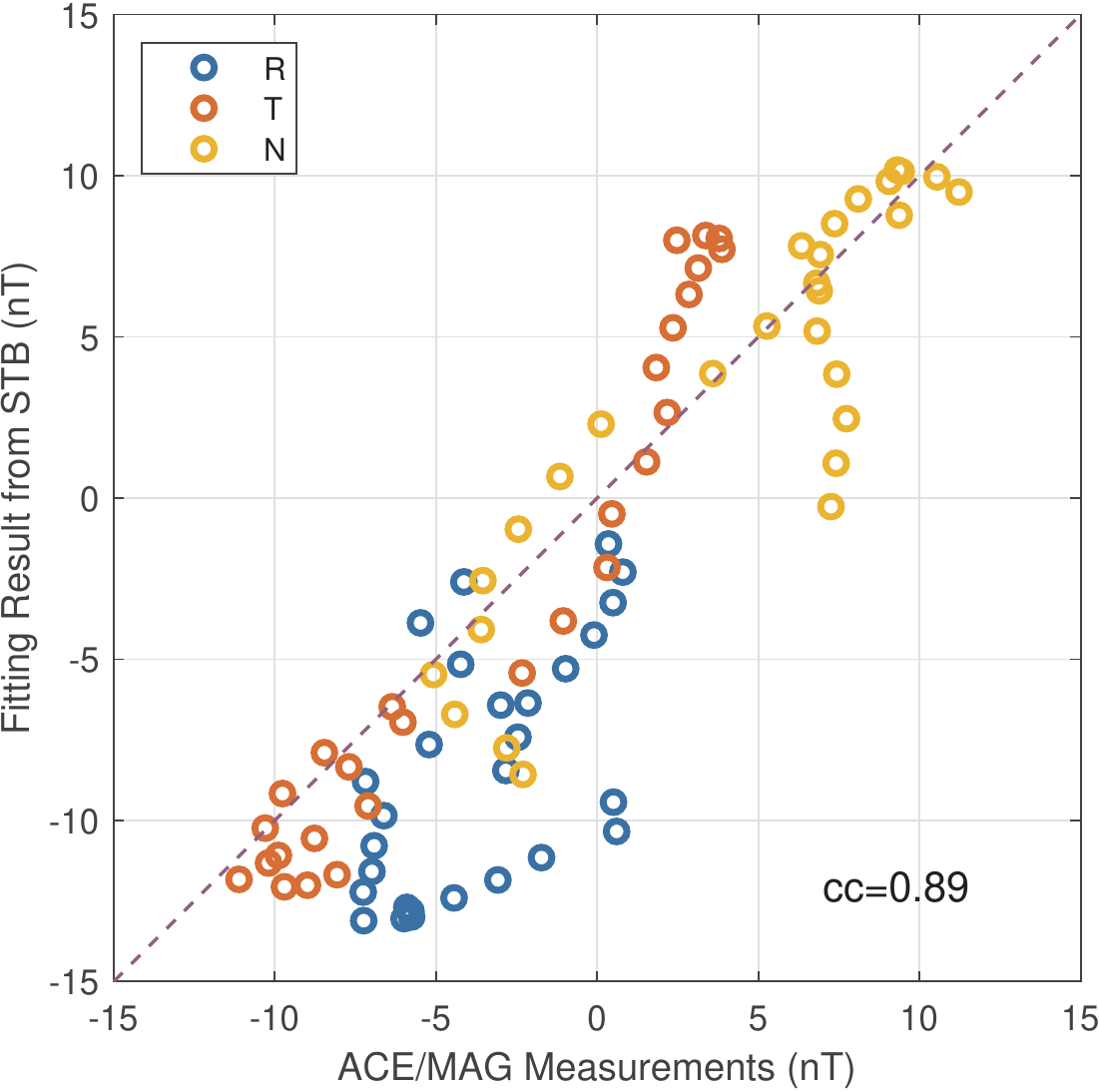}}\caption{Left panel: the spacecraft measurements with uncertainties (errorbars)  compared with the Freidberg solution (smooth solid curves) derived from the STB data, along the ACE spacecraft path. Magnetic field magnitude is in black. Right panel: The corresponding scatter plot of the magnetic field components between the two sets. The dashed line is a diagonal guideline. A  correlation coefficient $cc=0.89$ is calculated and denoted. }\label{fig:corr}
 \end{figure}
As a useful cross check, the corresponding {\em in situ} measurements along the ACE spacecraft path are shown in Figure~\ref{fig:ACEdata} (left). The magnetic field components within the MC interval are compared with the result obtained from the Freidberg solution based on the fitting of the STB data, presented earlier, i.e., along the green dots in Figure~\ref{fig:Bz0}. A quantitative point-by-point comparison and correlation are given in Figure~\ref{fig:corr}. The largest deviation between the analytic Freidberg solution and the actual ACE measurements results from the R component, whereas the other two components yield improved agreement \textbf{compared with the R component}, especially for the central portion of the interval. The corresponding correlation coefficient between the two entire sets is 0.89. 


\section{Summary and Discussion}
In summary, we have demonstrated, in detail, the practical procedures developed for the least-squares fitting of the Freidberg solution to {\em in situ} spacecraft measurements of MCs. The procedures 
are based, in part, on author's own experience with the GS reconstruction method as applied to the same type of structures in the solar wind, and 
strictly follow the $\chi^2$ minimization methodology \citep{2002nrca.book.....P}. The root-mean-squares estimates derived from higher resolution magnetic field vector measurements are taken as uncertainty estimates ($\sigma$) for the $\chi^2$ minimization formulation.  The method was applied to the two-spacecraft measurements of an MC in May 2007, when the STEREO  B spacescraft was separated from Earth by $\sim$ 3$^\circ$ longitudinally  in the HEE coordinates. Based on the {\em in situ} data and consistent with prior studies, STB and ACE spacecraft crossed the same MC structure along different paths, while the observational signatures at STA are more obscured during the times 18:00 UT 21 May - 06:00 UT 22 May 2007  (\textbf{see Figure~\ref{fig:ACEdata}, right panels, and also \citet{2009JGRAM}}). \textbf{It is likely because STA  missed the main flux rope structure as seen in Figure~\ref{fig:fldlines}.} The fitting result from STB data (with minimum $\chi^2=0.761$, and $Q=0.922$) yields a quasi-3D flux rope configuration, showing a winding flux bundle rooted on one major magnetic polarity region with right-handed chirality. The total axial magnetic flux content within the closed boundary defined by $B_z=0$ is $2.6\times10^{20}$ Mx. These results are consistent with \citet{2009JGRAM}, obtained from the GS reconstruction method. The ACE spacecraft was also crossing the same structure but along a separate path to the west of STB, as viewed toward the Sun. Comparison between the Freidberg solution based on the optimal fitting of STB measurements and the actual measured magnetic field components along the ACE spacecraft path yields a correlation coefficient $cc=$0.89. 

\begin{figure} 
 \centerline{\includegraphics[width=0.50\textwidth,clip=]{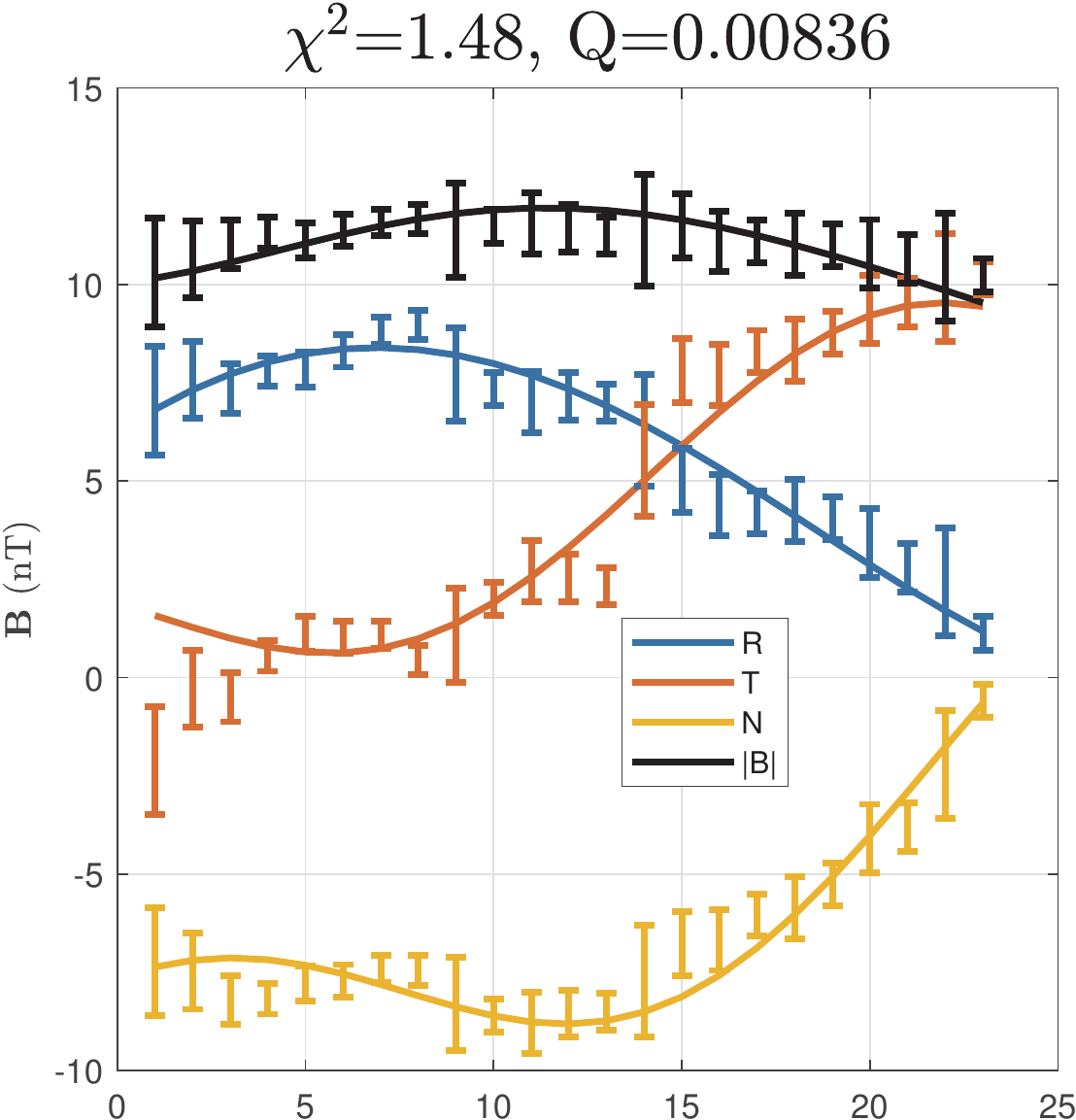}
  \includegraphics[width=0.4\textwidth,clip=]{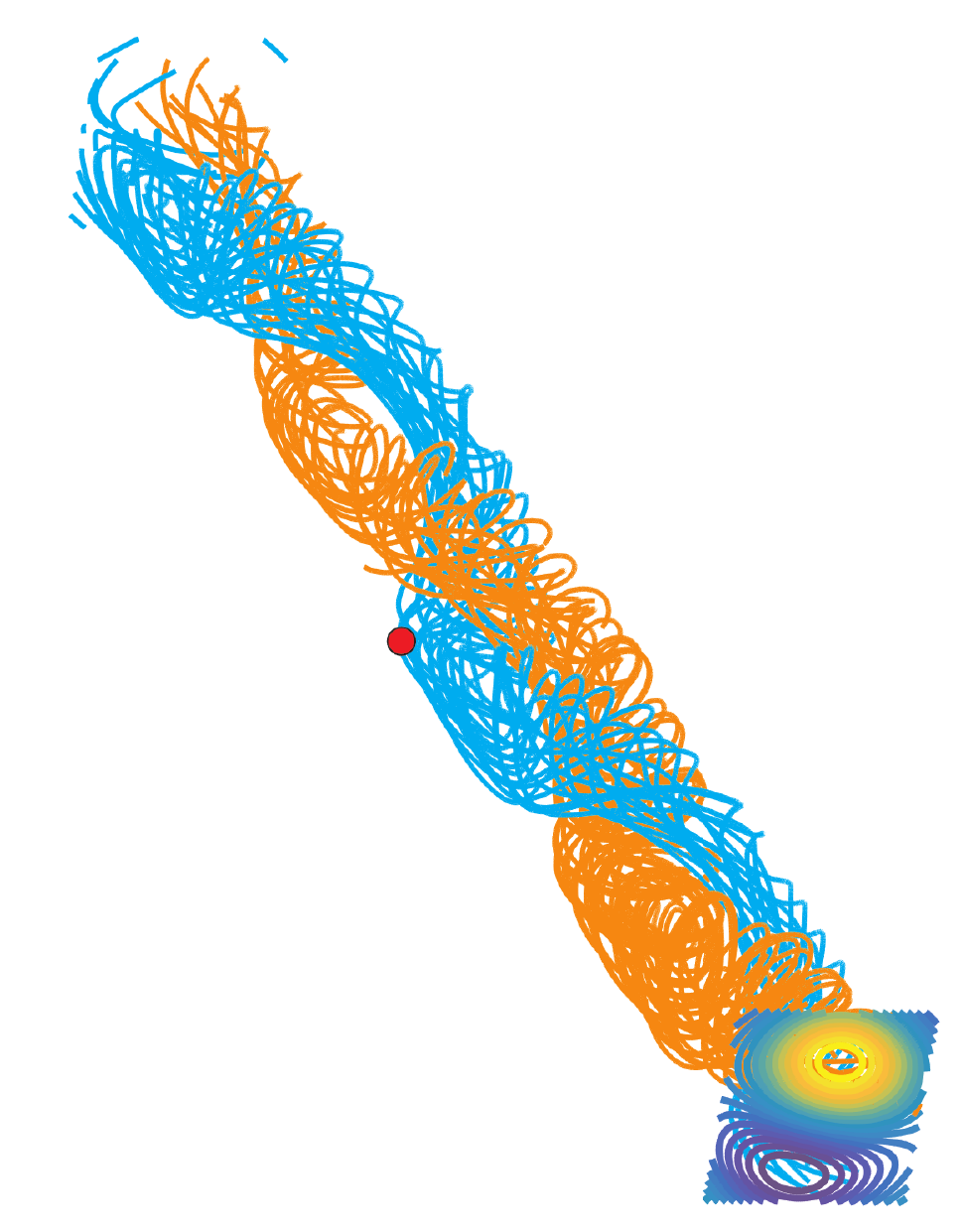}}
  \caption{\textbf{Left panel: the Freidberg solution for the STA MC interval marked in Figure~\ref{fig:ACEdata}, right panel; format is the same as Figure~\ref{fig:Brtn}. Right panel: the corresponding 3D view toward the Sun of the field-line configuration with  $\hat{\mathbf{z}}=[-0.8037,  -0.3613,  0.4727]$ in RTN coordinates; format is the same as Figure~\ref{fig:fldlines}. The STA spacecraft path is marked as red dots.}}\label{fig:STA}
 \end{figure}
We also applied the optimal fitting procedures described here to the other MC event on the 23 May 2007 examined by \citet{2009SoPhM}. \textbf{The event interval is marked in Figure~\ref{fig:ACEdata}, right panels. The corresponding fitting results are shown in Figure~\ref{fig:STA}.} We found an optimal fitting solution from the STA \emph{in situ} measurements with the minimum $\chi^2\approx 1.5$   and $Q\approx 0.0084$, which are marginal.  The reliability of this result is not as certain as the case presented in Section~\ref{sec:validation}. {\textbf{As seen from Figure~\ref{fig:STA}, right panel, and Figure~\ref{fig:STBdata} (left panel),  both the ACE and STB spacecraft  located to the east (left) of STA (the red dot). Therefore it resulted that both spacecraft did not cross the main body of the structure, based on the fitting result from STA}}. However in \citet{2009SoPhM}, a two-hour interval at the Wind spacecraft was compared favorably with the result from the GS reconstruction based on STA data. A threshold condition on the acceptable minimum $\chi^2$ values has yet to be further established  by more future event studies, in comparison with the corresponding GS reconstruction result and preferably combined with quantitative comparisons with solar source region properties, when available.

The merit of the Freidberg solution, despite its simple theoretical basis, is its representation of the complexity of an MC magnetic field toplogy at a higher level, for example, \textbf{by adding additional variability to the situation of mixed magnetic polarities as encompassed in the traditional Lundquist solution}. Being intrinsically 3D, it offers greater variabilities to the existing 1D and 2D models. However it remains a challenge, for all models, to satisfy all additional constraints presented by all available spacecraft measurements. As demonstrated by the case study of the rare occasion of two spacecraft with appropriate separation distance, the  spatial configuration of the MC structure was accounted for, by the Freidberg solution, to a good degree (as indicated by the minimum $\chi^2$ value from STB, and the correlation coefficient $cc$), but there is certainly room for improvement. Therefore we deem this new development to be a complementary tool to the existing MC analysis methods, especially the GS reconstruction method. Working together, they will enhance the existing capability of quantitatively characterizing the MC magnetic field topology based on {\em in situ} spacecraft measurements. 

%
\appendix   
\section{The de Hoffmann-Teller (HT) Analysis}\label{sec:app}
The de Hoffmann-Teller (HT) analysis is to determine an HT frame from the time-series data, following the approach given by \citet{1998ISSIRK}. One advantage of an HT frame is that by definition, in the HT frame, the electric field vanishes, then it follows from the Faraday's law, the time-dependence of the magnetic induction $\mathbf{B}$ should also vanish. Therefore the magnetic  field can be regarded to be stationary in time, consistent with the MC model assumptions.

Namely, for a time-series interval containing magnetic field $\mathbf{B}^{(m)}$ and plasma bulk flow velocity $\mathbf{V}^{(m)}$ measurements ($m=1,2, ..., M$) in the spacecraft frame, a constant HT frame velocity $\mathbf{V}_{HT}$ is obtained by minimizing \citep{1998ISSIRK}:
\begin{equation}
D_{HT}=\frac{1}{M}\sum_{m=1}^M|(\mathbf{V}^{(m)}-\mathbf{V}_{HT})\times \mathbf{B}^{(m)}|^2.\label{eq:HT}
\end{equation}
The quality of an HT frame is demonstrated by the component-wise plots of $\mathbf{E}_c=\mathbf{V}\times\mathbf{B}$ versus $\mathbf{E}_{HT}=\mathbf{V}_{HT}\times\mathbf{B}$, and $\mathbf{v'}=\mathbf{V}-\mathbf{V}_{HT}$ versus the Alfv\'en velocity $\mathbf{V}_A$. The latter (so-called Wal\'en plot) also indicates the relative importance of the inertia force (i.e., $\rho\mathbf{v'}\cdot\nabla\mathbf{v'}$) compared to the Lorentz force in the HT frame. Two metrics, the correlation coefficient  for the former, and the slope of a linear regression line for the latter, are calculated, respectively. 

\begin{figure}
 \centerline{\includegraphics[width=0.50\textwidth,clip=]{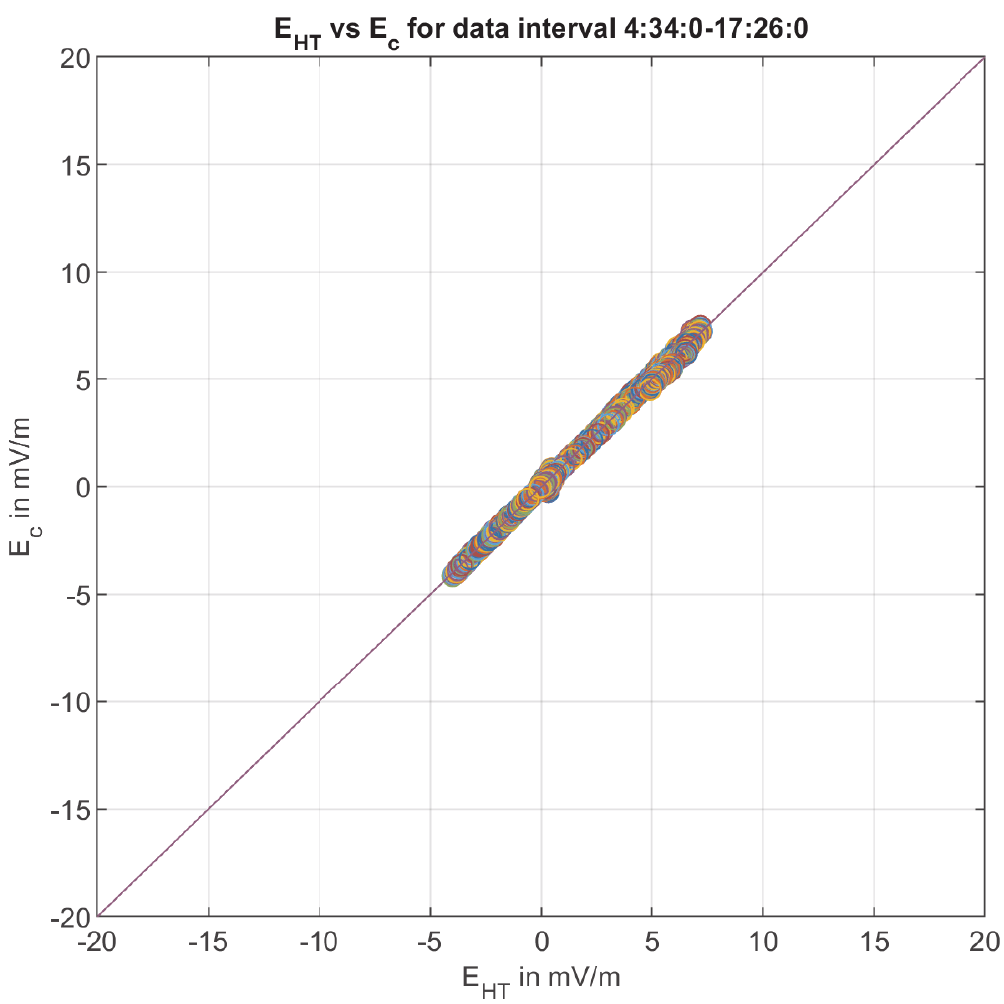}
  \includegraphics[width=0.5\textwidth,clip=]{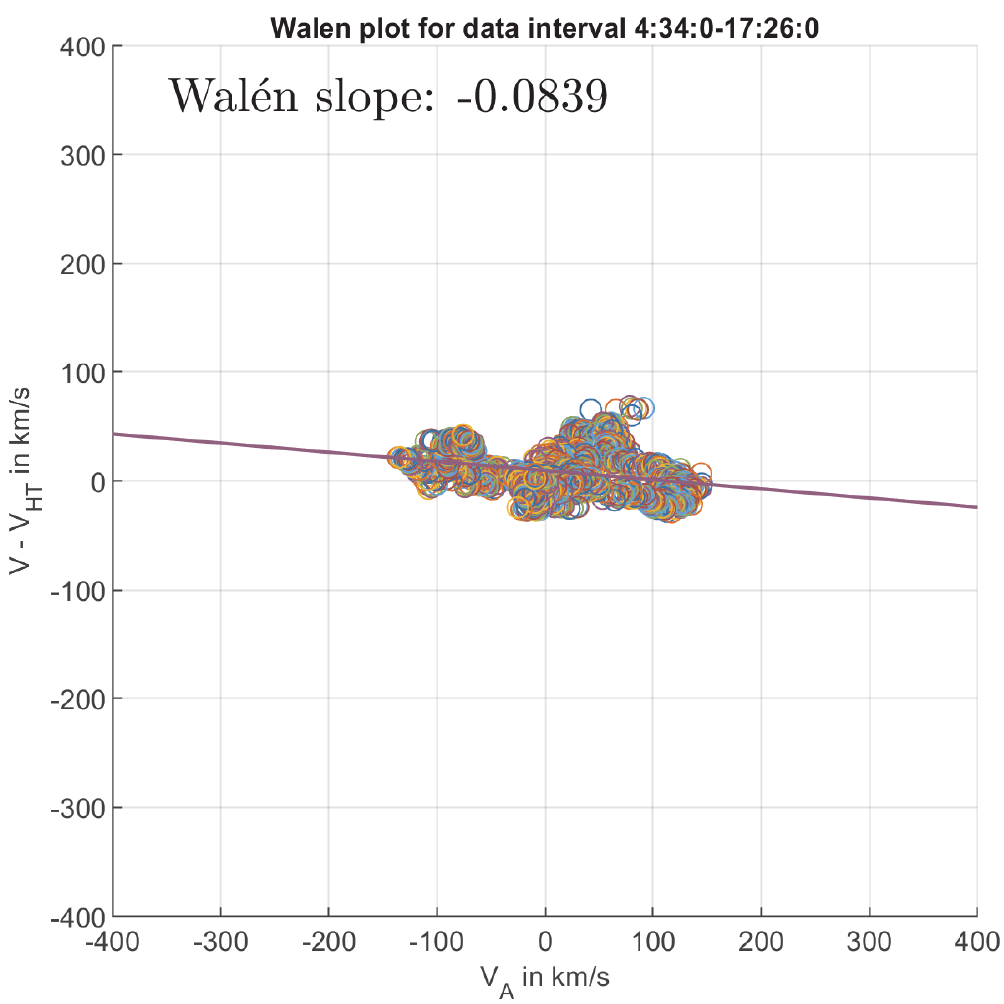}}
  \caption{The HT analysis results for the STB MC interval. Left panel: the component-wise plot of $\mathbf{E}_{HT}$ versus $\mathbf{E}_c$. Right panel: the component-wise plot of $\mathbf{v'}$ versus the Alfv\'en velocity in km/s. A linear regress line is drawn with the slope denoted in the top left corner. }\label{fig:HT}
 \end{figure}
For the STB MC interval given in Figure~\ref{fig:STBdata}, Figure~\ref{fig:HT} shows the HT analysis results  with the HT frame velocity $\mathbf{V}_{HT}=[ 440.16, -36.54, -0.08099]$ km/s in RTN coordinates. The  corresponding correlation coefficient and regression line slope are 
0.9990, and  -0.084, respectively. 
For the STA MC interval marked in Figure~\ref{fig:ACEdata} (right), the corresponding correlation coefficient and regression line slope are 0.9978, and  0.033, respectively, with $\mathbf{V}_{HT}= [482.66, 38,47, -16.28]$ km/s in RTN coordinates. Therefore for both cases, the assumption of time-stationary quasi-static equilibrium is satisfied when the analysis was carried out in the corresponding HT frame (with the correlation coefficient $\approx 1.0$ and the magnitude of Wal\'en slope $\ll 1.0$).

 \begin{acks}
  QH acknowledges NASA grants  80NSSC17K0016, 80NSSC18K0622,
80NSSC19K0276,  80NSSC21K0003, and NSF grants AGS-1650854, and AGS-1954503 for  support.  The ACE spacecraft Level2 data are accessed via the ACE Science Center (\url{http://www.srl.caltech.edu/ACE/ASC/}). The STEREO spacecraft data are accessed via the STEREO Science Center (\url{https://stereo-ssc.nascom.nasa.gov/}) and NASA CDAWeb (\url{https://cdaweb.gsfc.nasa.gov/index.html/}).
 \end{acks}

%
%
 \bibliographystyle{spr-mp-sola}
 \bibliography{ref_master3}  
%
%
%
%

\end{article} 
\end{document}